# EXPLOITING IMAGE LOCAL AND NONLOCAL CONSISTENCY FOR MIXED GAUSSIAN-IMPULSE NOISE REMOVAL

*Jian Zhang*[a], *Ruiqin Xiong*[b], *Chen Zhao*[b], *Siwei Ma*[b], *and Debin Zhao*[a]

[a]School of Computer Science and Technology, Harbin Institute of Technology, Harbin, China
[b]Institute of Digital Media, Peking University, Beijing, China

## ABSTRACT

Most existing image denoising algorithms can only deal with a single type of noise, which violates the fact that the noisy observed images in practice are often suffered from more than one type of noise during the process of acquisition and transmission. In this paper, we propose a new variational algorithm for mixed Gaussian-impulse noise removal by exploiting image local consistency and nonlocal consistency simultaneously. Specifically, the local consistency is measured by a hyper-Laplace prior, enforcing the local smoothness of images, while the nonlocal consistency is measured by three-dimensional sparsity of similar blocks, enforcing the nonlocal self-similarity of natural images. Moreover, a Split-Bregman based technique is developed to solve the above optimization problem efficiently. Extensive experiments for mixed Gaussian plus impulse noise show that significant performance improvements over the current state-of-the-art schemes have been achieved, which substantiates the effectiveness of the proposed algorithm.

## 1. INTRODUCTION

Image denoising is highly demanded in the field of image processing, since noise is usually inevitable during the process of image acquisition and transmission, which significantly degrades the image visual quality and increases the difficulty in the high-level image analysis [1-4].

There exist two different types of noise that are commonly encountered in real world: additive Gaussian noise and impulse noise. In literatures, there are numerous denoising methods that have been proposed separately for restoring images corrupted by either impulse noise or Gaussian noise. Here gives a brief review on the two types of noise, respectively.

Additive Gaussian noise is usually generated during image acquisition and characterized by adding each image pixel a value from a zero-mean Gaussian distribution. It is utilized to model thermal noise, and under certain conditions it is also the limit of other noises, such as photon counting noise and film grain noise. For Gaussian noise removal, we refer readers to [4] for a comprehensive review on the developments of additive Gaussian noise removal methods. It is important to stress that sparsity-based and non-local schemes have emerged as promising approaches with very impressive denoising results for Gaussian noise [12-14].

Impulse noise is often introduced by malfunctioning pixels in camera sensors, faulty memory locations in hardware, or transmission in a noisy channel, which can be classified into two categories, namely salt-and-pepper noise and random-valued impulse noise [20]. Specifically, for images corrupted by impulse noise, the intensity values of corrupted pixels are replaced with either the extreme value or random numbers in the dynamic range of images, while the remainders are left unchanged. There are mainly two types of methods for the restoration of images corrupted by impulse noise. The first type is median filter that is widely used for its denoising ability and computational efficiency [5, 8, 9]. In order to better preserve the edge structures of images, the other type of variational approaches have been developed for impulse noise removal [18, 3, 10, 11]. In [18], a data-fidelity term of $l^1$ norm was first introduced to achieve a significant improvement for impulse noise removal. In order to resolve this problem, many effective two-phase methods are proposed [3, 7, 19] by associating various variational models with different median filters. The first phase of their methods is to detect the location of noisy pixels corrupted by impulse noise using median filters, and then employ some variational methods to estimate the gray values for the noisy pixels in the second phase.

As a matter of fact, we often encounter the case where an image is corrupted by both Gaussian and impulse noise in practice. Such mixed noise could occur when an image that has already been contaminated by Gaussian noise in the procedure of image acquisition with faulty equipment suffers impulsive corruption during its transmission over noisy channels successively.

However, not much work has been designed to effectively eliminate mixed noise due to the distinct characteristics of both types of degradation processes. Concretely, the methods developed for Gaussian noise cannot effectively suppress impulse noise because they interpret the noisy pixels as edges to be preserved, whereas the approaches for impulse noise removal will retain most Gaussian noise in the restored images leading to grainy, visually disappointing results [18].

The mainstream algorithms are developed to combine variational models with different median filters based on the idea of above two-phase scheme. More precisely, by extending [18] to restrict the data fidelity term within the outlier domain, Cai et al. proposed a modified two-phase method to deblur images corrupted by impulse noise plus Gaussian noise [7]. In [10], a total variation prior based model was also designed for impulse and Gaussian noise removal. Lately, Li et al. further proposed to handle blurred images corrupted by mixed Gaussian-impulse noise by minimizing a new functional including a content-dependent fidelity term and a novel regularizer defined as the $l^1$ norm of geometric tight framelet coefficients [11].

Although the regularization terms above have the edge-preserving property, they only consider image local information and tend to generate over-smoothed results. Hence, the more accurate and effective image prior is desirable. Moreover, the characteristic of self-similarity in natural images is not utilized.

In this paper, a new variational algorithm for mixed Gaussian-impulse noise removal is proposed within regularization framework. Our main contributions are two-fold. First, a generalized variational scheme for mixed Gaussian-impulse noise removal is formulated via exploiting image local consistency and nonlocal consistency simultaneously. Second, a Split-Bregman based iterative numerical algorithm is developed to solve the above optimization problem efficiently.

The rest of this paper is organized as follows. In Section 2, a generic problem formulation for mixed Gaussian-impulse noise removal within regularization framework is given. We introduce the image local and nonlocal consistency in Section 3. The details of our proposed algorithm are presented in Section 4. Experiments are reported in Section 5 and we conclude this paper in Section 6.

## 2. PROBLEM FORMULATION

As a fundamental problem in the field of image processing, image restoration aims to reconstruct the original high-quality image from its degraded observed version. In this paper, we restrict our attention to the task of removing a mixed noise composed of Gaussian noise plus impulse noise. It is a typical ill-posed problem, which can be generally modeled as the following two steps:

$$\begin{cases} \tilde{y} = x + n \\ y = \mathbb{N}_{imp}(\tilde{y}) \end{cases}, \quad (1)$$

where $x, y, n \in \mathbf{R}^N$ are lexicographically stacked representations, denoting the original image, the degraded image and the additive Gaussian white noise with standard variance $\sigma$, respectively. $\mathbb{N}_{imp}$ represents the corruption by impulse noise with a corruption rate $r \in [0,1]$.

This Gaussian noise process will typically result from the physical limitations of the image acquisition procedure: thermal noise, photon counting noise and film grain noise [9], as explained before.

There are two common types of impulse noise used in a wide variety of applications: salt-and-pepper and random-valued impulse noise [3]. Denote by $x[i]$ the intensity value of the $i$-th pixel in image $x$ and by $[d_{min}, d_{max}]$ the dynamic range of an image. Then the two models of impulse noise are defined by

a.  Salt-and-pepper noise:

$$y[i] = \begin{cases} d_{min}, & \text{with probability } r/2, \\ d_{max}, & \text{with probability } r/2, \\ \tilde{y}[i], & \text{with probability } 1\text{-}r, \end{cases}$$

where $r$ determines the level of the salt-and-pepper noise.

b.  Random-valued impulse noise:

$$y[i] = \begin{cases} d_i, & \text{with probability } r, \\ \tilde{y}[i], & \text{with probability } 1\text{-}r, \end{cases}$$

where $d_i$ are identically and uniformly distributed random numbers in $[d_{min}, d_{max}]$ and $r$ determines the level of the random-valued impulse noise.

Let $\mathcal{N}$ be the set of locations of the outlier candidate pixels corrupted by impulse noise and $\mathcal{B} = \mathcal{A} / \mathcal{N}$ be the set of locations of the left pixels without impulse noise. Denote $y_\mathcal{B}, x_\mathcal{B}, n_\mathcal{B}$ as the vectors of elements of $y, x, n$, respectively, whose locations belong to $\mathcal{B}$ and denote $y_\mathcal{N}$ as the vector of elements of $y$, whose locations belong to $\mathcal{N}$, that is, $y_\mathcal{N} = y / y_\mathcal{B}$.

Assume the cardinality of $\mathcal{B}$ is M and denote $B$ as an $M \times N$ matrix of indicators $\{0,1\}$ showing which elements of $y$ belong to $\mathcal{B}$, so that

$$\begin{cases} y_\mathcal{B} = B \cdot y \\ x_\mathcal{B} = B \cdot x \\ n_\mathcal{B} = B \cdot n \end{cases}. \quad (2)$$

Because the impulse corrupted pixels located in $\mathcal{N}$ do not carry any information about the original image or the previous Gaussian noise, the problem of removing mixed Gaussian-impulse noise can be converted to the problem of inferring the complete noiseless data $x$ from its partial Gaussian noisy data $y_\mathcal{B}$. To determine $B$, like [10, 11], we use adaptive median filter (AMF) [8] for salt-and-pepper noise detection and adaptive center-weighted median filter (ACWMF) [9] for random-valued impulse noise detection since they are simple and effective. Thus, when given $B$, the generic variational model for mixed Gaussian-impulse noise removal can be formulated as follows:

$$\hat{x} = \underset{x}{\operatorname{argmin}} \|B \cdot x - y_\mathcal{B}\|_2^2 + \lambda \cdot \varphi(x), \quad (3)$$

where the parameter $\lambda$ plays the role of balancing the data fidelity term and the regularization term.

## 3. IMAGE LOCAL AND NONLOCAL CONSISTENCY

In this section, we give the definitions of image local and nonlocal consistency, which will be incorporated into the variational framework as regularizations for mixed noise removal.

### 3.1. Image Local Consistency

From the view of statistics, the image is preferred when its responses for a set of filters are as small as possible. In most of the best-performing methods, the marginal statistics are assumed to be Laplacian. However, studies of real-world images have indicated that the marginal distributions have significantly heavier tails than Laplacian, being well modeled by a hyper-Laplacian form. Hyper-Laplacian image priors have been exploited in a wide range of settings [22-24]: image deblurring, super-resolution, transparency separation, which have achieved better quality results.

In this paper, we measure the image local consistency by hyper-Laplacian priors, defined by

$$\Phi_{LC}(\boldsymbol{x}) = \|\nabla \boldsymbol{x}\|^{2/3}, \quad (4)$$

where $\nabla$ represents the spatial gradients of the image in both vertical and horizontal directions. The image local consistency essentially enforces smoothness between neighboring pixels of natural images in a statistical manner.

### 3.2. Image Nonlocal Consistency

Inspired by the success of sparse representation [13] and self-similarity [12] in image restoration [14], we integrate them and introduce a type of nonlocal three-dimensional sparsity as a measurement of image nonlocal consistency, which can be formulated in the following four steps:

Firstly, divide the image $\boldsymbol{x}$ with size $N$ into $n$ overlapped blocks of size $B_s$ and each block is denoted by $\boldsymbol{x}_k$, i.e., $k=1,2,...,n$. Secondly, define $S_{\boldsymbol{x}_k}$ the set including the $c$ best matched blocks to $\boldsymbol{x}_k$ in the $L \times L$ training window, that is, $S_{\boldsymbol{x}_k} = \{S_{\boldsymbol{x}_k \otimes 1}, S_{\boldsymbol{x}_k \otimes 2},...,S_{\boldsymbol{x}_k \otimes c}\}$. Thirdly, for every $S_{\boldsymbol{x}_k}$, a group is formed by stacking the blocks belonging to $S_{\boldsymbol{x}_k}$ into a three-dimensional array, which is denoted by $Z_{\boldsymbol{x}_k}$. Finally, denote $T^{3D}$ the operator of a three-dimensional transform, and $T^{3D}(Z_{\boldsymbol{x}_k})$ the transform coefficients for $Z_{\boldsymbol{x}_k}$. Let $\Theta_{\boldsymbol{x}}$ be the column vector with size $K = B_s \cdot c \cdot n$ built from all the $T^{3D}(Z_{\boldsymbol{x}_k})$ arranged in lexicographic order. Therefore, the image nonlocal consistency can be measured by the nonlocal three-dimensional sparsity of $\Theta_{\boldsymbol{x}}$, written as

$$\Phi_{NC}(\boldsymbol{x}) = \|\Theta_{\boldsymbol{x}}\|_0 = \sum_{k=1}^{n} \|T^{3D}(Z_{\boldsymbol{x}_k})\|_0, \quad (5)$$

where $\|*\|_0$ is $l^0$ norm, counting the nonzero entries of a vector.

Similarly, the inverse operator $\Omega_{NC}$ corresponding to $\Phi_{NC}$ can be defined in the reverse procedures. Thus, given $\Theta_{\boldsymbol{x}}$, the new estimate of $\boldsymbol{x}$ is expressed as $\hat{\boldsymbol{x}} = \Omega_{NC}(\Theta_{\boldsymbol{x}})$.

## 4. AN ITERATIVE ALGORITHM FOR MIXED GAUSSIAN-IMPULSE NOISE REMOVAL

By incorporating image local consistency (4) and nonlocal consistency (5) into the generic variational model (3), a new formulation for mixed Gaussian-impulse noise removal can be expressed as follows:

$$\hat{\boldsymbol{x}} = \arg\min_{\boldsymbol{x}} \|\boldsymbol{B} \cdot \boldsymbol{x} - \boldsymbol{y}_{\mathcal{B}}\|_2^2 + \lambda \Phi_{LC}(\boldsymbol{x}) + \beta \Phi_{NC}(\boldsymbol{x}), \quad (6)$$

where $\lambda$ and $\beta$ are control parameters.

Problem (6) is essentially non-convex and quite difficult to solve directly due to the non-differentiability and non-linearity of the two consistency terms. Solving it efficiently is one of the main contributions of this paper.

By utilizing variable splitting technique [21], the problem will change into an equivalent constrained optimization:

$$(\hat{\boldsymbol{x}}, \hat{\boldsymbol{u}}, \hat{\boldsymbol{w}}) = \underset{\boldsymbol{x},\boldsymbol{u},\boldsymbol{w}}{\operatorname{argmin}} \|\boldsymbol{B} \cdot \boldsymbol{x} - \boldsymbol{y}_{\mathcal{B}}\|_2^2 + \lambda \Phi_{LC}(\boldsymbol{u}) \\ + \beta \Phi_{NC}(\boldsymbol{w}) \quad s.t. \quad \boldsymbol{x}=\boldsymbol{u}, \boldsymbol{x}=\boldsymbol{w} \quad (7)$$

Applying Bregman algorithm [16, 17] to (7) leads to the following iterative steps:

$$(\hat{\boldsymbol{x}}^{(j+1)}, \hat{\boldsymbol{u}}^{(j+1)}, \hat{\boldsymbol{w}}^{(j+1)}) = \underset{\boldsymbol{x},\boldsymbol{u},\boldsymbol{w}}{\operatorname{argmin}} \|\boldsymbol{B} \cdot \boldsymbol{x} - \boldsymbol{y}_{\mathcal{B}}\|_2^2 + \lambda \|\nabla \boldsymbol{u}\|^{2/3} \\ + \beta \|\Theta_{\boldsymbol{w}}\|_0 + \mu_1 \|\boldsymbol{x} - \boldsymbol{u} - \boldsymbol{b}^{(j)}\|_2^2 + \mu_2 \|\boldsymbol{x} - \boldsymbol{w} - \boldsymbol{c}^{(j)}\|_2^2 \quad (8)$$

$$\boldsymbol{b}^{(j+1)} = \boldsymbol{b}^{(j)} - (\hat{\boldsymbol{x}}^{(j+1)} - \hat{\boldsymbol{u}}^{(j+1)}); \\ \boldsymbol{c}^{(j+1)} = \boldsymbol{c}^{(j)} - (\hat{\boldsymbol{x}}^{(j+1)} - \hat{\boldsymbol{w}}^{(j+1)}). \quad (9)$$

Instead of solving (8) directly, here, an alternating direction technique is employed, which alternatively minimizes one variable while fixing the other variables, to split Problem (8) into the following three sub-problems. In what follows, we argue that every separated sub-problem admits a closed form solution. For simplicity, the subscript $j$ is omitted without confusion.

### 4.1. $\boldsymbol{x}$ sub-problem

Given $\boldsymbol{u}$ and $\boldsymbol{w}$, $\boldsymbol{x}$ sub-problem becomes

$$\min_{\boldsymbol{x}} \|\boldsymbol{B} \cdot \boldsymbol{x} - \boldsymbol{y}_{\mathcal{B}}\|_2^2 + \mu_1 \|\boldsymbol{x} - \boldsymbol{u} - \boldsymbol{b}\|_2^2 + \mu_2 \|\boldsymbol{x} - \boldsymbol{w} - \boldsymbol{c}\|_2^2. \quad (10)$$

Since (10) is a minimization problem of strictly convex quadratic function, there is a closed form, expressed as

$$\hat{\boldsymbol{x}} = (\boldsymbol{B}^T \boldsymbol{B} + \mu \boldsymbol{I})^{-1} \cdot \boldsymbol{s}. \quad (11)$$

Here, $\boldsymbol{s} = \boldsymbol{B}^T \boldsymbol{y} + \mu_1(\boldsymbol{b}+\boldsymbol{u}) + \mu_2(\boldsymbol{c}+\boldsymbol{w})$, $\boldsymbol{I}$ is identity matrix and $\mu = \mu_1 + \mu_2$. Owing to the particular structure of Matrix $\boldsymbol{B}$ that satisfies $\boldsymbol{B}^T \boldsymbol{B} = \boldsymbol{I}$, applying the Sherman-Morrison-Woodbury matrix inversion formula to (11) yields

$$\hat{\boldsymbol{x}} = \frac{1}{\mu}\left(\boldsymbol{I} - \frac{1}{1+\mu} \boldsymbol{B}^T \boldsymbol{B}\right) \cdot \boldsymbol{s}. \quad (12)$$

### 4.2. $u$ sub-problem

With the recovered $x$ and $w$, $u$ sub-problem is

$$\min_{u} \mu_1 \|u-(x-b)\|_2^2 + \lambda \|\nabla u\|^{2/3}. \quad (13)$$

According to [22], the solution to (13) can be obtained analytically by just solving a quadratic function.

### 4.3. $w$ sub-problem

Given $x$ and $u$, we get $w$ sub-problem

$$\min_{w} \beta \|\Theta_w\|_0 + \mu_2 \|x-w-c\|_2^2$$
$$= \min_{w} \left\{ \frac{1}{2}\|w-r\|_2^2 + \frac{\beta}{2\mu_2} \sum_{k=1}^{n} \|T^{3D}(Z_{w_k})\|_0 \right\}, \quad (14)$$

where $r=(x-c)$. Here, we model the elements of $w-r$ as random variables from a Gaussian process with zero mean and variance $\sigma^2$, which is reasonable and commonly used in practice. Under this assumption, since $w, r \in \mathbf{R}^N, \Theta_w, \Theta_r \in \mathbf{R}^K$, and the transform $T^{3D}$ is orthogonal for every group, there exist the following two equations with very large probability (limited to 1):

$$\|w-r\|_2^2 / N = \sigma^2, \quad (15)$$

$$\|\Theta_w - \Theta_r\|_2^2 / K = \sum_{k=1}^{n} \|T^{3D}(Z_{w_k}) - T^{3D}(Z_{r_k})\|_2^2 / K = \sigma^2 \quad (16)$$

Incorporating (15) and (16) into (14) leads to

$$\min_{w} \frac{1}{2} \|\Theta_w - \Theta_r\|_2^2 + \frac{K\alpha}{N\theta} \|\Theta_w\|_0. \quad (17)$$

Owing to [21] the closed form of (17) is written as $\tilde{\Theta}_w = \text{hard}(\Theta_r, \sqrt{2\tau})$, where $\tau = K\alpha/N\theta$, $\cdot$ stands for the element-wise product of two vectors and $\text{hard}(\cdot, a)$ is hard thresholding function with threshold $a$. Thus, the solution for the $w$ sub-problem (14) is

$$\hat{w} = \Omega_{NC} \tilde{\Theta}_w = \Omega_{NC}\left(\text{hard}(\Theta_r, \sqrt{2\tau})\right). \quad (18)$$

**Table 1.** The Proposed Algorithm for Mixed Gaussian-Impulse Noise Removal via Image Local and Nonlocal Consistency

**Input**: Degraded image $y$ corrupted by Gaussian plus impulse noise and the standard variance $\sigma$ of Gaussian noise $n$.
**Initialization:**
$\hat{x}^{(0)} = \hat{u}^{(0)} = \hat{w}^{(0)} = y, b^{(0)} = c^{(0)} = 0, \mathcal{M}^{(0)} = \varnothing$;
**Outer Loop:** Iterate on $i = 0, 1, 2, 3$
 Apply AMF (for salt and pepper noise) or ACWMF (for random-valued noise) with the threshold $\delta^{(i+1)}$ to the image $\hat{x}^{(i)}$ to get the noisy candidate set $\mathcal{M}^{(i+1)}$;
 Let $\mathcal{N} = \mathcal{M}^{(i)} \cup \mathcal{M}^{(i+1)}$ and build sub-sampling matrix $B$;
 **Inner Loop:** Iterate on $j = 0, 1, 2, \ldots, t$
  Solve $x$ sub-problem to achieve $\hat{x}$ by computing Eq. (11);
  Solve $u$ sub-problem to achieve $\hat{u}$ by computing Eq. (13);
  Solve $w$ sub-problem to achieve $\hat{w}$ by computing Eq. (18);
  Update $b$ and $c$ by computing Eq. (9);
**Output**: Final restored image $\hat{x}$.

### 4.4. Summary of the Proposed Algorithm

To further improve the quality of denoising results, the progressive mechanism for identifying the noisy candidate set is introduced [19]. Hence, the proposed algorithm for mixed Gaussian-impulse noise removal via image local and nonlocal consistency is summarized in Table 1.

## 5. EXPERIMENTAL RESULTS

In this section, extensive experimental results are provided to evaluate the performance of the proposed algorithm. In the simulations, images will be corrupted by Gaussian noise with standard deviation $\sigma$ and impulse noise density level $r$. Note that the density level can be detected automatically by the median filters [8, 9], thus the only parameter required known is standard deviation $\sigma$ of Gaussian noise. Two state-of-the-art algorithms compared with our proposed method are: TV [10], IFASDA [11].

Extensive experiments are carried out on four benchmark images, where the standard variance $\sigma$ of Gaussian noise equals 10 and the impulse noise level $r$ varies from 30% to 50% for salt-and-pepper noise and from 10% to 30% for random-valued noise.

Table 2 and Table 3 present the PSNR results of the three comparative denoising algorithms on all test images for Gaussian plus salt-and-pepper impulse noise and Gaussian plus random-valued impulse noise, respectively. Obviously, the proposed method considerably outperforms the other methods in all the cases, with a PSNR improvement of about 2 dB on average over the second best algorithm (i.e. IFASDA [11]). In particular, for Image *Barbara*, which is rich in textures, in the case of Gaussian plus salt-and-pepper impulse noise with $\sigma = 10$ and $r = 50\%$, the PSNR gain achieved by the proposed method over IFASDA is as high as 3.6 dB.

Some visual results of the recovered images for the three algorithms are presented in Figures 1~4. **Please enlarge and view the figures on the screen for better comparison.** One can see that TV [10] is effective in suppressing the noises; however, it produces over-smoothed results and eliminates much image details (see Fig. 1(b), Fig. 2(b), Fig. 3(b) and Fig. 4(b)). IFASDA is very competitive in recovering the image structures. However, it tends to generate some annoying artifacts in the smooth regions (see Fig.1(c), Fig. 2(c), Fig. 3(c) and Fig. 4(c)). By comparing with TV and IFASDA, the proposed method not only successfully suppresses the mixed Gaussian-impulse noise but also reconstructs quite accurate textures and preserves more and sharper image details, exhibiting the best visual quality consistently (see Fig. 1(d), Fig. 2(d), Fig. 3(d) and Fig. 4(d)). The high performance of the proposed algorithm is attributed to the employment of image local and nonlocal consistency at the same time, which offers a powerful mechanism of characterizing the statistical properties of natural images.

**Table 2.** PSNR of Various Methods for Gaussian plus Salt-and-Pepper Noise Removal (dB)

| Image | Lena | | | Boat | | | House | | | Barbara | | | Avg. |
|---|---|---|---|---|---|---|---|---|---|---|---|---|---|
| r (%) | 30 | 40 | 50 | 30 | 40 | 50 | 30 | 40 | 50 | 30 | 40 | 50 | |
| Noisy | 10.63 | 9.40 | 8.44 | 10.66 | 9.42 | 8.46 | 10.69 | 9.46 | 8.50 | 10.58 | 9.36 | 8.39 | 9.50 |
| TV | 31.33 | 30.85 | 30.20 | 29.20 | 28.53 | 27.66 | 31.63 | 31.10 | 30.36 | 26.86 | 26.18 | 25.40 | 29.11 |
| IFASDA | 32.69 | 32.27 | 31.70 | 30.82 | 30.28 | 29.50 | 32.68 | 32.26 | 31.69 | 29.47 | 28.59 | 27.45 | 30.78 |
| Proposed | **34.02** | **33.59** | **33.00** | **31.54** | **30.95** | **30.17** | **34.86** | **34.43** | **33.82** | **32.33** | **31.95** | **31.12** | **32.65** |

**Table 3.** PSNR of Various Methods for Gaussian plus Random-Valued Impulse Noise Removal (dB)

| Image | Lena | | | Boat | | | House | | | Barbara | | | Avg. |
|---|---|---|---|---|---|---|---|---|---|---|---|---|---|
| r (%) | 10 | 20 | 30 | 10 | 20 | 30 | 10 | 20 | 30 | 10 | 20 | 30 | |
| Noisy | 18.78 | 16.05 | 14.36 | 18.88 | 16.13 | 14.42 | 18.87 | 16.08 | 14.42 | 18.65 | 15.93 | 14.24 | 16.40 |
| TV | 31.52 | 30.94 | 30.16 | 28.83 | 28.10 | 27.39 | 31.54 | 30.89 | 29.84 | 25.47 | 24.88 | 24.18 | 28.65 |
| IFASDA | 32.17 | 31.47 | 30.43 | 29.29 | 28.46 | 27.64 | 31.88 | 31.13 | 29.87 | 25.69 | 25.06 | 24.30 | 28.95 |
| Proposed | **33.63** | **32.72** | **31.76** | **31.04** | **29.64** | **28.80** | **34.14** | **33.30** | **32.20** | **30.80** | **29.20** | **27.62** | **31.24** |

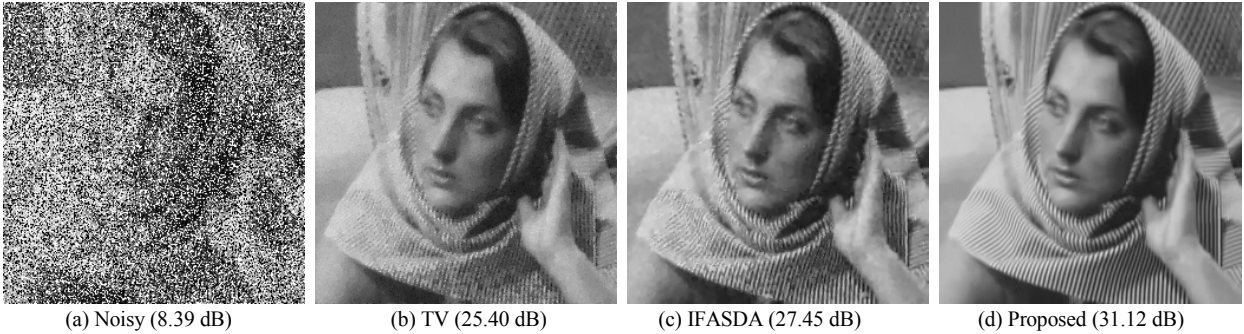

(a) Noisy (8.39 dB)    (b) TV (25.40 dB)    (c) IFASDA (27.45 dB)    (d) Proposed (31.12 dB)

**Fig. 1.** Denoised results of various methods on Image *Barbara* corrupted by Gaussian plus salt-and-pepper impulse noise with $\sigma = 10$ and $r = 50\%$.

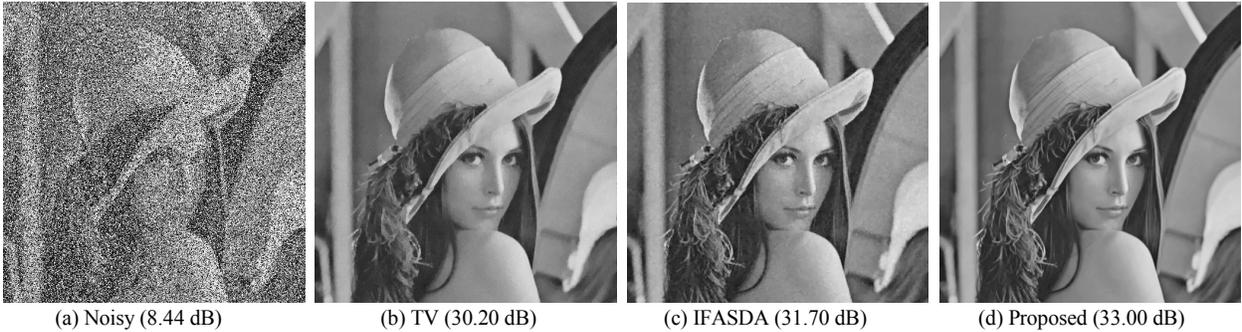

(a) Noisy (8.44 dB)    (b) TV (30.20 dB)    (c) IFASDA (31.70 dB)    (d) Proposed (33.00 dB)

**Fig. 2.** Denoised results of various methods on Image *Lena* corrupted by Gaussian plus salt-and-pepper impulse noise with $\sigma = 10$ and $r = 50\%$.

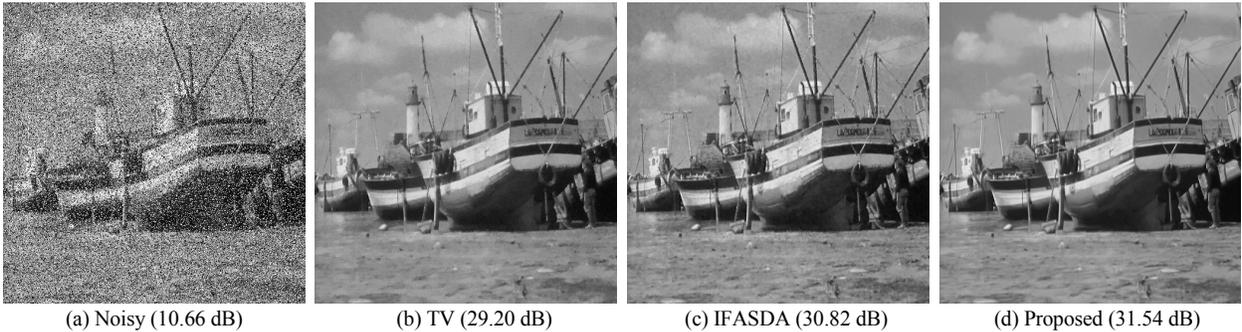

(a) Noisy (10.66 dB)    (b) TV (29.20 dB)    (c) IFASDA (30.82 dB)    (d) Proposed (31.54 dB)

**Fig. 3.** Denoised results of various methods on Image *Boat* corrupted by Gaussian plus salt-and-pepper impulse noise with $\sigma = 10$ and $r = 30\%$.

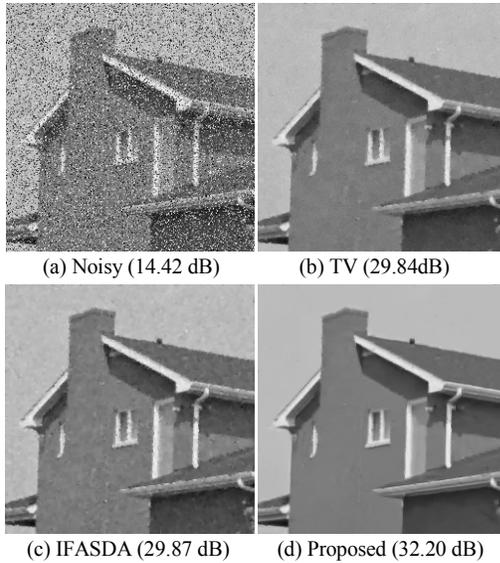

(a) Noisy (14.42 dB)  (b) TV (29.84dB)
(c) IFASDA (29.87 dB)  (d) Proposed (32.20 dB)

**Fig. 4.** Denoised results of various methods on Image *House* corrupted by Gaussian plus random-valued impulse noise with $\sigma = 10$ and $r = 30\%$.

## 6. CONCLUSIONS

In this paper, a novel algorithm for mixed Gaussian-impulse noise removal is proposed by exploiting image local consistency and nonlocal consistency simultaneously, which can efficiently characterize the statistical properties of natural images. Extensive experiments demonstrate that the proposed algorithm is able to achieve significant performance improvements over the current state-of-the-art schemes, generating denoising results with better quality. Current and future work includes the extensions on a variety of applications, such as deblurring images corrupted by mixed Gaussian-impulse noise.

## 7. ACKNOWLEDGEMENT

We would like to thank Dr. Lixin Shen of [11] for kindly providing his code. This work was supported in part by the Major State Basic Research Development Program of China (2009CB320905), National Science Foundation of China (No. 61073083 and No. 61100096).